\documentclass[aps,a4paper,12pt,onecolumn,notitlepage,nobibnotes,nofootinbib,
superscriptaddress,noshowpacs,centertags,tightenlines]{revtex4}
\begin{document}
\selectlanguage{english}

\title{DETECTABILITY OF DARK MATTER SUBHALOS BY MEANS OF THE GAMMA-400 TELESCOPE}   

\author{\firstname{A.~E.}~\surname{Egorov}}
\email{eae14@yandex.ru}
\affiliation{Lebedev Physical Institute, RU-119991 Moscow, Russia}
\author{\firstname{A.~M.}~\surname{Galper}}
\affiliation{Lebedev Physical Institute, RU-119991 Moscow, Russia}
\affiliation{National Research Nuclear University MEPhI, RU-115409 Moscow,Russia}
\author{\firstname{N.~P.}~\surname{Topchiev}}
\affiliation{Lebedev Physical Institute, RU-119991 Moscow, Russia}
\author{\firstname{A.~A.}~\surname{Leonov}}
\affiliation{Lebedev Physical Institute, RU-119991 Moscow, Russia}
\affiliation{National Research Nuclear University MEPhI, RU-115409 Moscow,Russia}
\author{\firstname{S.~I.}~\surname{Suchkov}}
\affiliation{Lebedev Physical Institute, RU-119991 Moscow, Russia}
\author{\firstname{M.~D.}~\surname{Kheymits}}
\affiliation{National Research Nuclear University MEPhI, RU-115409 Moscow,Russia}
\author{\firstname{Yu.~T.}~\surname{Yurkin}}
\affiliation{National Research Nuclear University MEPhI, RU-115409 Moscow,Russia}

\begin{abstract}
We investigated the detectability of Galactic subhalos with masses $(10^6-10^9)M_{\odot}$ formed by annihilating WIMP dark matter by the planned GAMMA-400 gamma-ray telescope. The inner structure of dark matter subhalos and their distribution in the Galaxy were taken from corresponding simulations. We showed that the expected gamma-ray flux from subhalos strongly depends on WIMP mass and subhalo concentration, but less strongly depends on the subhalo mass. In an optimistic case we may expect the flux of 10--100 ph/year above 100 MeV from the closest and most massive subhalos, which would be detectable sources for GAMMA-400. However, resolving the inner structure of subhalos might be possible only by the joint analysis of the future GAMMA-400 data and data from other telescopes due to smallness of fluxes. Also we considered the recent subhalo candidates 3FGL J2212.5+0703 and J1924.8--1034 within the framework of our model. We concluded that it is very unlikely that these sources belong to the subhalo population.
\end{abstract}

\maketitle

\section{Introduction and motivation}\label{intro}

Dark matter (DM) in the Universe was discovered over 80 years ago. However, its exact physical nature remains to be one of the biggest puzzles in the modern astrophysics. There is a big variety of candidates on the role of DM. Weakly Interacting Massive Particles (WIMPs) in the form of neutralino probably still remain to be the most popular candidate (see, e.g., \cite{Bertone-book}). As known, these supersymmetric particles can self-annihilate with production of highly energetic Standard Model particles: photons, electrons, protons, etc. \cite{Bertone-book}. This process is actively going in any regions with high DM concentration and could manifest itself by an emission from annihilation products. This is very promising direction of DM searches (so-called indirect searches), which are being conducted in any astrophysical objects from the Sun to galaxy clusters. One such an object of interest is DM subhalos in our Galaxy. N-body simulations of DM halos reliably predict an existence of extended hierarchical substructure inside them in the form of subhalos in a very wide mass range down to $\sim 10^{-6} M_{\odot}$ \cite{Bertone-book}. Thus, the modern simulations (see, e.g., \cite{2016PhR...636....1C}) of Milky Way (MW) --sized halos are able to resolve tens of thousands of subhalos. The most massive subhalos are expected to host the known dwarf satellites of MW. Smaller subhalos (with masses less than $(10^7-10^8) M_{\odot}$ \cite{PhysRevD.96.063009}) may completely lack any stellar/baryonic component and, hence, any astrophysical counterpart on the sky. However, they could be close enough in order to represent bright gamma-ray sources due to DM annihilation. At the same time, $\sim 1/3$ of the sources in the 3FGL Fermi-LAT catalog lack firm associations with known counterparts at other wavelengths \cite{2015ApJS..218...23A}. Hence, some of these unidentified sources could be, in fact, DM subhalos. In general, one may expect the following valuable properties of DM subhalos as gamma-ray sources, which allow one to classify them unambiguously \cite{2016PhR...636....1C}:
\begin{itemize}
\item they almost completely lack any emission other than that from DM annihilation;
\item the latter is absolutely stable emission with no variability in time;
\item all subhalos would have the same predictable spectra;
\item they would have almost isotropic distribution on the sky;
\item they might be resolved as extended sources (!), which is rather impossible for typical astrophysical sources, such as blazars.
\end{itemize}

Thus, DM subhalos represent very exciting targets for indirect DM searches. Naturally, their detectability by Fermi-LAT has been already widely explored (e.g., \cite{PhysRevD.96.063009}). The conclusion of this exploration was not optimistic: Fermi-LAT is able to detect $\sim 1$ such object on the whole sky and resolve spatially <1 object. At the same time, some groups made attempts to identify DM subhalos in Fermi-LAT point source catalogs (e.g., \cite{2016JCAP...05..049B}). These searches yielded a couple of interesting candidates 3FGL J2212.5+0703 and J1924.8--1034 \cite{2016JCAP...05..049B, 2016PhRvD..94l3002W, 2017PhRvD..95j2001X}. Their exact nature remains uncertain, and we will discuss it in Section \ref{J}. 

Also, it is very timely and relevant now to evaluate a potential of future gamma-ray missions to detect and resolve subhalos (e.g., \cite{2017arXiv170908562C}). The main goal of our paper is to explore the detectability of DM subhalos by the new gamma-ray telescope GAMMA-400, which is under designing now and planned for launch around 2025 \cite{2013AdSpR..51..297G, 2017EPJWC.14506001T}. It is expected to have unique characteristics, such as energy and angular resolutions. The latter is designed to be $\sim 0.01\degree$ at energies $\sim$ 100 GeV, which is by an order of magnitude better than that for Fermi-LAT. Hence, GAMMA-400 may have a great potential to detect DM subhalos and resolve their inner structure, paving the road towards a comprehensive exploration of DM nature. Also we may expect a fruitful outcome from the joint analysis of GAMMA-400 observational data together with data from other telescopes - Fermi-LAT, proposed e-ASTROGAM \cite{2017ExA...tmp...24D} and others. We perform our analysis assuming that DM is constituted by the mentioned neutralinos.

\section{Modeling the subhalo population}\label{model}

Here we used the quite comprehensive model of Galactic subhalos developed in \cite{PhysRevD.96.063009}. The authors of this work performed N-body simulations of the MW-like halo of two types: with cold DM only and with CDM+baryons. The authors did not notice significant differences in the results of these two cases. The mass range of subhalos, which are resolvable in this simulation, is $\sim (10^6-10^{11})M_{\odot}$. In our work we focused only on the range $(10^6-10^9)M_{\odot}$, since according to \cite{PhysRevD.96.063009} all the heavier subhalos host stars and, hence, formally represent dwarf satellites. Indeed, the boundary between subhalos and dwarfs is very nominal - these objects share the same nature. However, we plan to study signals from dwarfs in a separate work. Concerning the masses $(10^6-10^9)M_{\odot}$, \cite{PhysRevD.96.063009} predicts $\sim 4000$ of such objects in MW assuming a quite standard mass distribution $dN/dM \sim M^{-1.9}$.

Let's briefly describe main properties of subhalos, which were inferred based on \cite{PhysRevD.96.063009}. The DM density distribution inside the subhalo is approximated by the Einasto profile:
\begin{equation}\label{eq-ro}
	\rho (r) = \rho_s \exp\left(-\frac{2}{\alpha}\left(\left(\frac{r}{r_s}\right)^{\alpha}-1\right)\right),
\end{equation}
where $r$ is the distance from the subhalo center, $r_s, \rho_s$ are the scale radius and density, $\alpha = 0.16$. Figures 2,10 in \cite{PhysRevD.96.063009} show the dependence of $r_s$ on the subhalo mass found in the simulation. One can see a large dispersion of radii values for each mass. It is important to study this radii uncertainty in details, since as we will see below, the gamma-ray fluxes from subhalos strongly depend on $r_s$ (mainly due to a quadratic dependence of the flux on $\rho (r)$ - see (\ref{eq-f})). For this purpose we decided to compare the simulation results with the mass distribution inside real MW satellites - dwarf galaxies. We used \cite{2015MNRAS.451.2524M, 2015ApJ...801...74G}, where the authors reproduced the mass distribution inside the known dwarfs employing the observational data on their kinematics. Under the mass distribution we assume essentially DM distribution, since the baryonic component is negligible. Our comparison showed that overall the simulation reproduces the real dwarfs with a reasonable accuracy. However, we noticed the following systematic trend - the real dwarfs typically have smaller $r_s$ than those from the simulation. If to plot the dwarfs in Figure 2 of \cite{PhysRevD.96.063009}, they would be located near the bottom boundary of the cloud of points from the simulation. In such a situation we made a decision to take $r_s-M$ dependence shown by the real dwarfs as a realistic (i.e., average expectation) scenario. Such $r_s$ correspond to the minimal values found in the simulation. This dependence was approximated by the following equation:
\begin{equation}\label{eq-rs}
	\lg(r_s/\mbox{kpc}) = 0.441\lg(M_{200}/M_{\odot})-4.10,
\end{equation}
where $M_{200}$ is the virial subhalo mass, i.e. the mass inside the sphere with average density 200$\rho_{crit}$. $r_s$ sets a value of the concentration parameter $c_{200} \approx r_{200}/r_s$, which strongly influences on the flux value from a subhalo. It is important to estimate the uncertainty range of final fluxes due to the uncertainties in $r_s$. For this purpose, we calculated the fluxes not only with the mean $r_s$ (or concentration) model (\ref{eq-rs}), but also with extreme models, which would approximately represent marginal cases, i.e., the minimal and maximal possible fluxes. We defined these marginal concentration models as follows. For the conservative end (i.e., the minimal concentration and flux) we took $r_s-M_{200}$ relation, which was obtained as the mean trend in the simulation \cite{PhysRevD.96.063009}. It is shown by the blue line in Figure 2 of \cite{PhysRevD.96.063009} and can be approximated by the following equation:
\begin{equation}\label{eq-rsh}
	\lg(r_s/\mbox{kpc}) = -0.111\lg^2(M_{200}/M_{\odot})+2.11\lg(M_{200}/M_{\odot})-10.0.
\end{equation}
Such $r_s$ values exceed (\ref{eq-rs}) by a factor of $\sim 2$ for all the masses. Quite naturally, as the lowest $r_s$ we decided to take 1/2 of the values defined by (\ref{eq-rs}). In fact, such a choice of the minimal/maximal $r_s$ is not very random: the resulting range of the allowed $r_s$ values approximately correspond to the range formed by the measurement uncertainties of $r_s$ in the real dwarfs according to \cite{2015MNRAS.451.2524M}.

Also we have to model the distance to the subhalos in order to calculate the flux from them. We extracted the distance--mass relation from Figure 10 of \cite{PhysRevD.96.063009}. Naturally we considered the subhalos closest to us. We approximated the distance to them by the following relation, which roughly provides $\sim 1$ subhalo with such a distance on the whole sky in each mass interval with the width of one order of magnitude: 
\begin{equation}\label{eq-d}
  \lg(d_{min}/\mbox{kpc}) = 0.0505\lg^2(M_{200}/M_{\odot})-0.556\lg(M_{200}/M_{\odot})+2.76.
\end{equation}
Thus, we may expect a few objects on the whole sky so nearby to us in the chosen mass range of $(10^6-10^9)M_{\odot}$. This monotonous relation gives $d_{min}(10^6 M_{\odot}) = 17$ kpc and $d_{min}(10^9 M_{\odot}) = 70$ kpc.

\section{The results - subhalo fluxes and sizes}\label{res}

The gamma-ray flux in the energy range $[E_{min}..E_{max}]$ inside the solid angle $\Delta \Omega$ can be calculated by the following well-known equation:
\begin{equation}\label{eq-f}
\begin{aligned}
F_{\gamma}(\Delta\Omega) =
    & \underbrace{ \frac{1}{4\pi} \frac{\langle \sigma v \rangle}{2m_{\chi}^{2}}\int^{E_{\max}}_{E_{\min}}\frac{dN_{\gamma}}{dE_{\gamma}}dE_{\gamma}}_{\mbox{\footnotesize{particle~physics}}} 
    & \times \underbrace{ \int_{\Delta\Omega}\int_{\rm l.o.s.}\rho^2(\vec{r})dl d\Omega'}_{\mbox{\footnotesize{J~factor}}},
\end{aligned}
\end{equation}
where $m_{\chi}$ is DM particle mass, $\langle \sigma v \rangle$ is its annihilation cross section and $dN_{\gamma}/dE_{\gamma}$ is the gamma-ray spectrum from one annihilation (modeled by the tools \cite{PPPC, 2011JCAP...03..051C, 2011JCAP...03..019C}). The spectra of the signal significantly depend on the annihilation channel, which is unknown a priori. In this aspect we followed a standard approach and considered the two representative annihilation channels: $\chi\chi \rightarrow b \bar{b}$, $\chi\chi \rightarrow \tau^+\tau^-$. Another non-trivial aspect of the flux computation is that the DM density $\rho(r)$ in (\ref{eq-f}) must generally include not only the smooth component (\ref{eq-ro}), but also the substructure component inside a subhalo, i.e. sub-substructure. The boost factor to the signal due to such a sub-substructure was studied in \cite{2017MNRAS.466.4974M} with the following conclusion: in the case of relevant subhalos the total signal is boosted by $\leqslant 10\%$ (see Figure 7 there). Our estimates are not so precise, therefore we do not take into account the sub-substructure effect here.
 
\begin{figure}[t]
\includegraphics[width=0.495\textwidth]{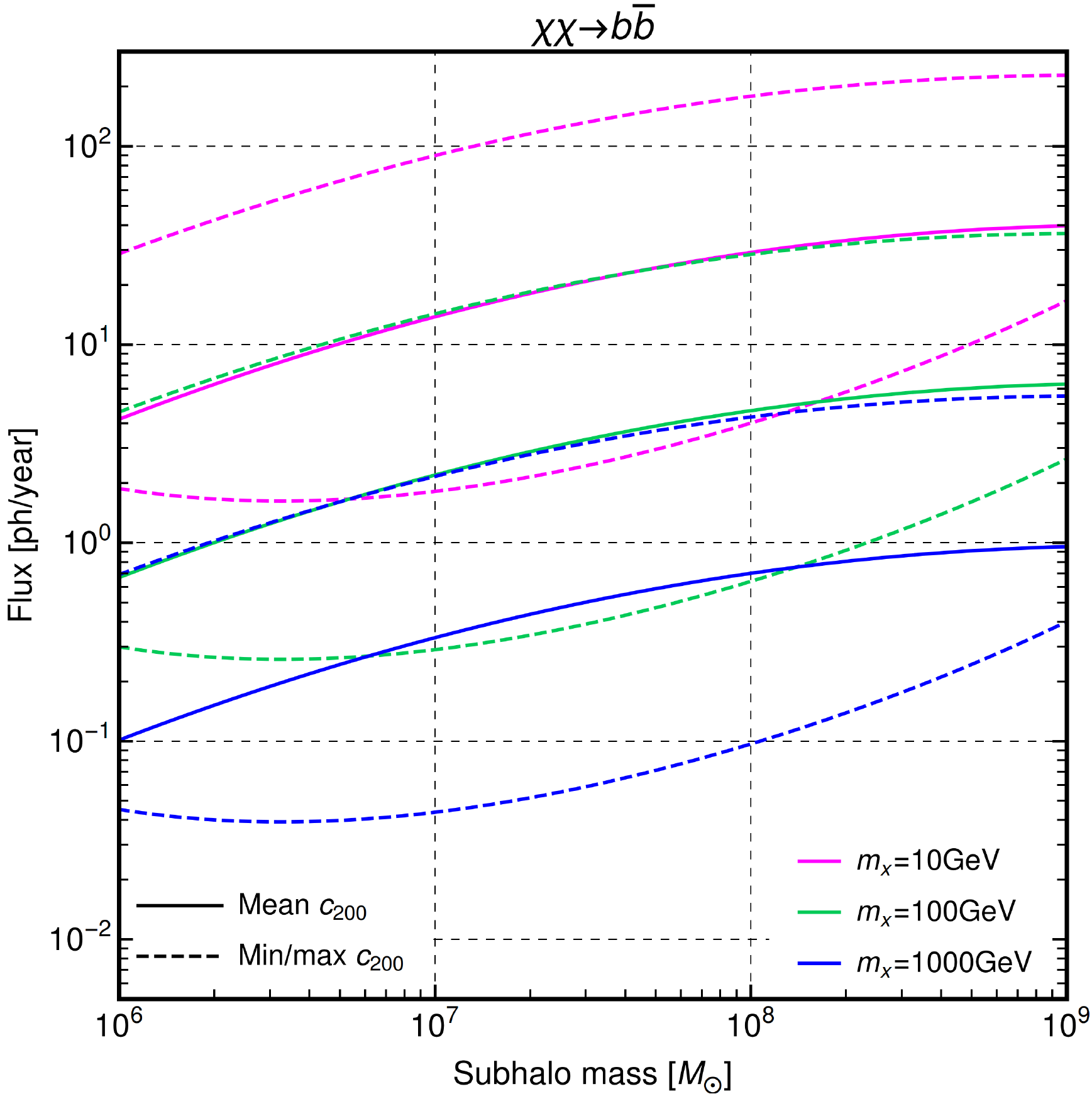}
\includegraphics[width=0.495\textwidth]{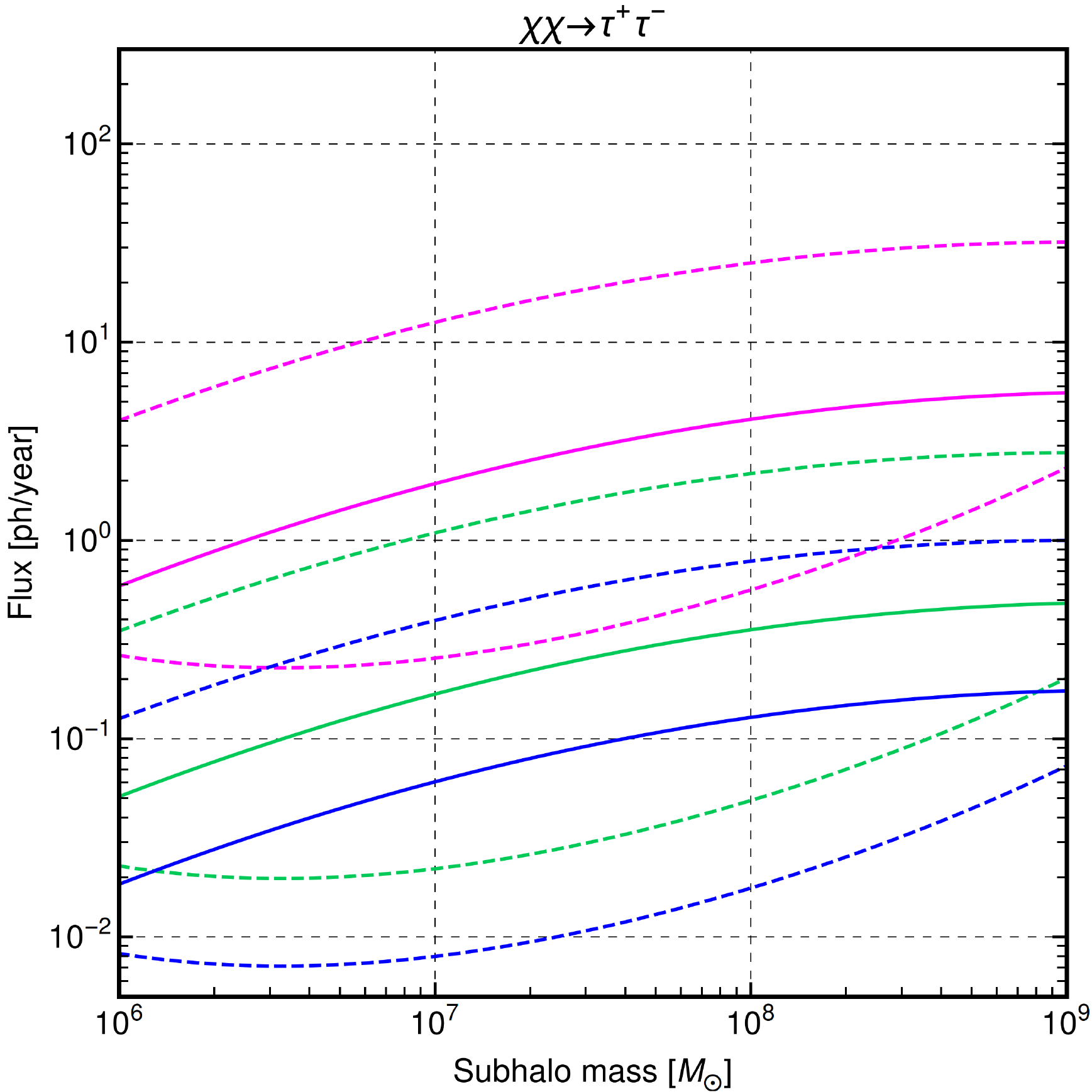}
\caption{Dependence of the gamma-ray flux above 100 MeV from the closest subhalos on their mass for different annihilation channels and WIMP masses. The solid lines correspond to the average (realistic) model of the subhalo concentration, while the dashed lines reflect the extreme cases of the concentration. The annihilation cross section for each WIMP mass corresponds to the Fermi-LAT upper limits by observations of dwarfs \cite{2015PhRvL.115w1301A} (see Table \ref{tab}). The effective detector area is equal to 4000 cm$^2$. For more details see Sections \ref{model},\ref{res}.}
\label{fig1}
\end{figure}

At this point we substituted Equations (\ref{eq-ro})-(\ref{eq-d}) into (\ref{eq-f}) and computed the final fluxes from subhalos. The results are shown in Figure \ref{fig1} ($E_{\gamma} > 100$ MeV) for various WIMP masses, annihilation channels and subhalo concentration models discussed in Section \ref{model}. We chose the annihilation cross section values for each mass according to the latest Fermi-LAT constraints by observations of 15 dwarfs \cite{2015PhRvL.115w1301A} (Figure 1 there) as the maximally allowed cross section. Specific values are presented in Table \ref{tab}. The fluxes are presented in the units of ph/year specifically for the GAMMA-400 gamma-ray telescope (normal incidence is assumed). However, one can easily convert those into the standard units of ph s$^{-1}$ cm$^{-2}$ dividing by the assumed telescope effective area $A = 4000~\mbox{cm}^2$ and number of seconds in a year. In Figure \ref{fig1} we can see that the flux dependence on subhalo mass is rather mild. The most massive subhalos are the brightest even being the remotest. Naively extrapolating such a trend to the masses below $10^6 M_{\odot}$ (which are not resolvable by the simulation), we may conclude that such small subhalos would not be bright to be well detectable. However, indeed one has to study the lower mass end properly to make any precise conclusions. Also, besides an individual detection, numerous low mass subhalos in principle may contribute to the angular power spectrum of gamma-ray background fluctuations.

Also from Figure \ref{fig1} we may note that the flux strongly depends on the assumed concentration model: the former varies by about an order of magnitude when the concentration is varied between the minimal and maximal cases discussed in Section \ref{model}. DM particle mass plays an important role too: the flux quickly decreases with the mass increase, even though the cross section increases (see Table \ref{tab}). Also one can see that the leptonic annihilation channel provides significantly lower integral fluxes in comparison with the hadronic one. Overall, for some parameter combinations we may expect the flux of 10--100 ph/year, which makes these objects detectable by GAMMA-400.

\begin{table}[h]
\caption{WIMP annihilation cross section values, which were chosen for all flux calculations. They correspond to the largest cross sections allowed by Fermi-LAT limits from observations of dwarfs \cite{2015PhRvL.115w1301A}. } \label{tab}
\bigskip
\begin{tabular}{|c|c|c|}
  \hline
    $m_{\chi}$, GeV & $\langle \sigma v \rangle (\chi\chi \rightarrow b \bar{b})$, cm$^3$/s & $\langle \sigma v \rangle (\chi\chi \rightarrow \tau^+\tau^-)$, cm$^3$/s \\
  \hline
  10 & $5 \cdot 10^{-27}$  & $4 \cdot 10^{-27}$ \\
  100 & $3 \cdot 10^{-26}$ & $3 \cdot 10^{-26}$ \\
  1000 & $2 \cdot 10^{-25}$ & $8 \cdot 10^{-25}$ \\[1mm]
  \hline
\end{tabular}
\end{table}

\begin{figure}[t]
\includegraphics[width=0.495\textwidth]{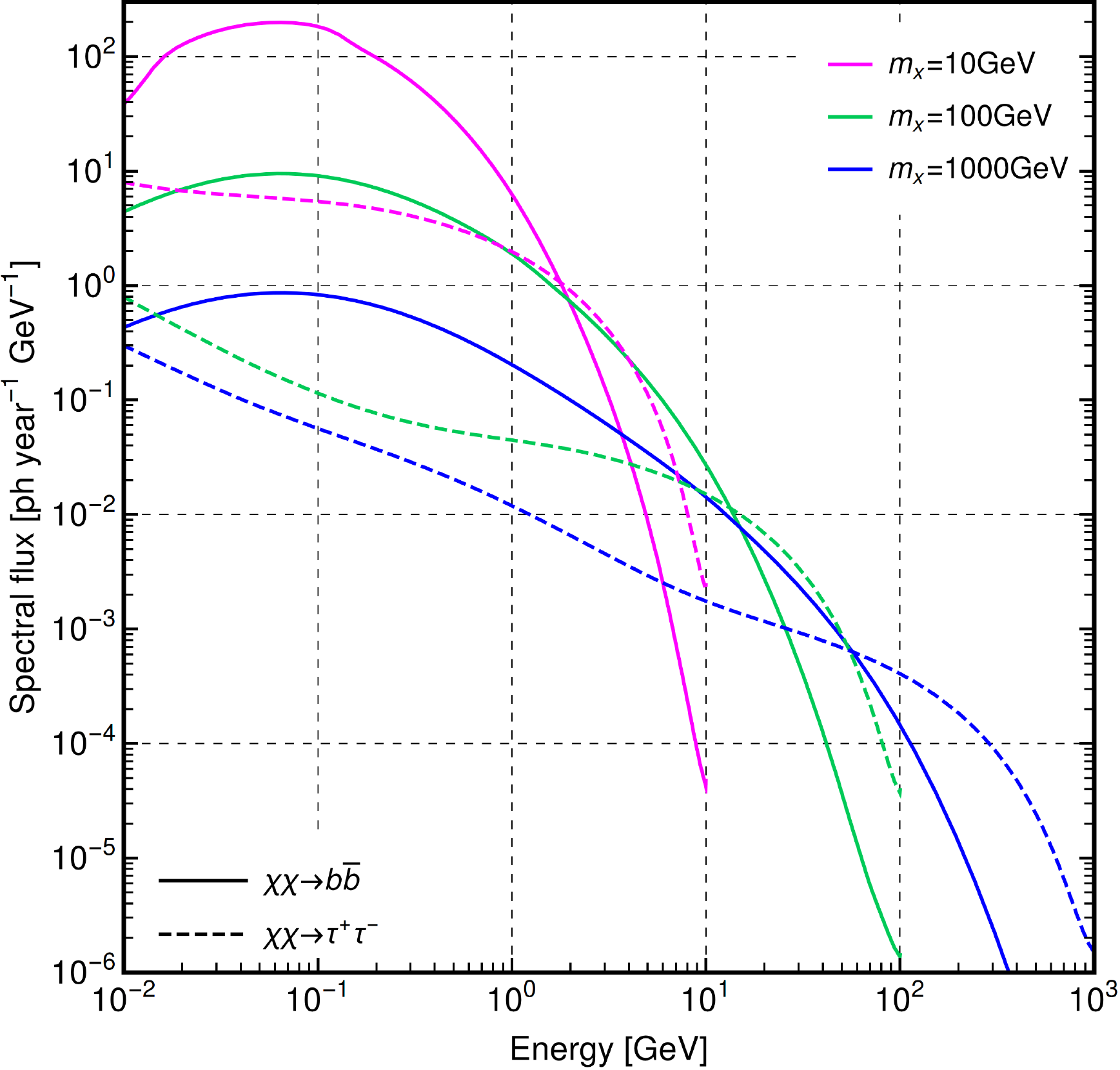}
\includegraphics[width=0.495\textwidth]{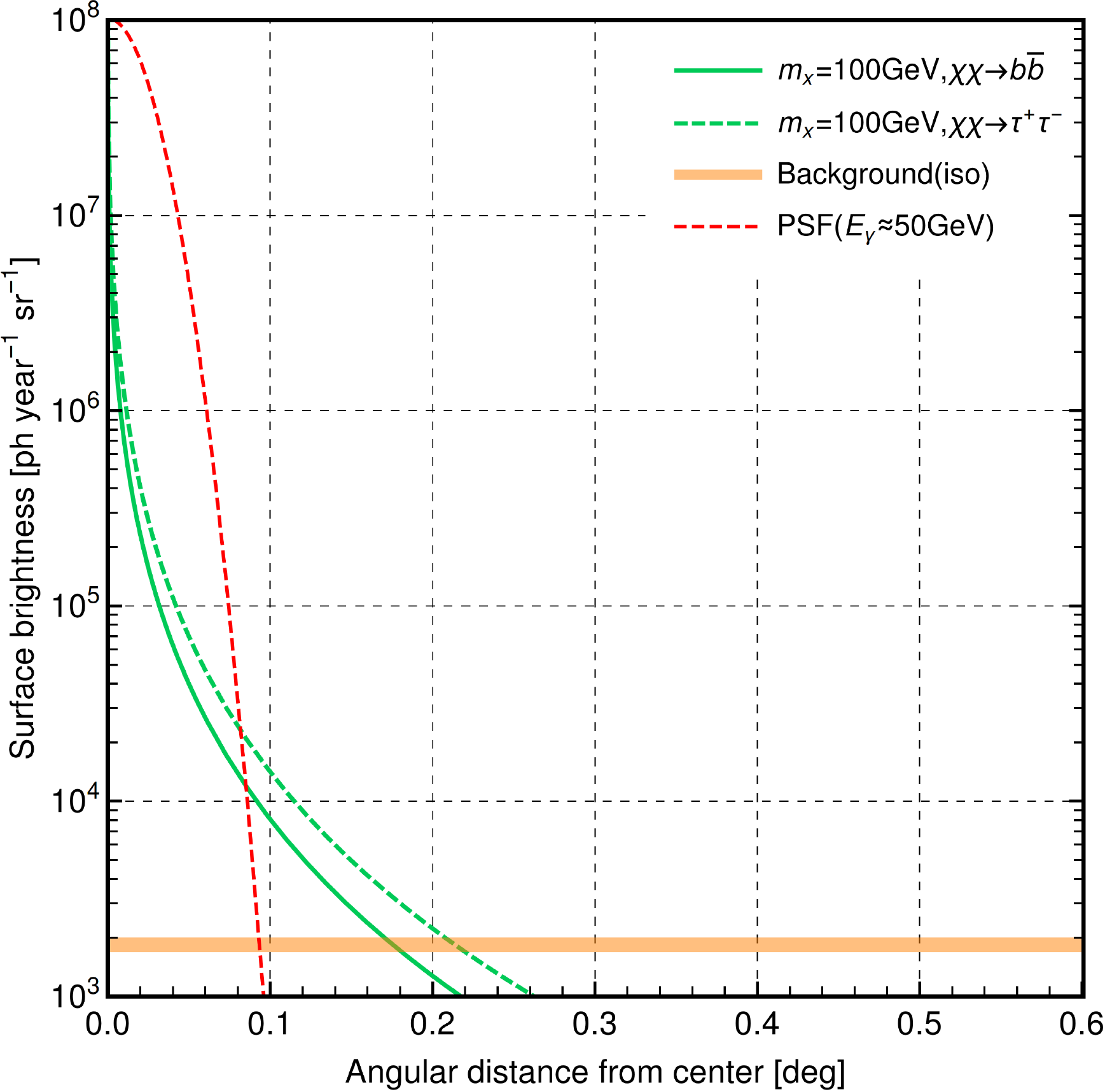}
\caption{\textit{Left:} The spectra of a nearby subhalo with mass $10^9 M_{\odot}$ for various WIMP masses and annihilation channels. The annihilation cross section is similar to that in Figure \ref{fig1}, the subhalo concentration model is medium (realistic). \textit{Right:} an example of the surface brightness distribution for the case of a nearby subhalo with mass $10^9 M_{\odot}$ in the energy range 10--100 GeV. The concentration model is maximal. Such a model provides the flux $\sim 1$ ph/year in the mentioned energy range. The red dashed line shows an approximate PSF of the instrument. The orange stripe shows an approximate brightness of the isotropic background according to Fermi-LAT data \cite{Fermi}. The right border of the plot  ($0.6\degree$) corresponds to $r \approx r_s$ for this subhalo. For more details see Sections \ref{model},\ref{res}. }
\label{fig2}
\end{figure}

Figure \ref{fig2}, left shows the spectra of a massive subhalo for various WIMP masses and annihilation channels. We may notice that at low energies energies the spectral flux from $b \bar{b}$ channel dominates over that from $\tau^+\tau^-$ and at higher energies ($E_{\gamma} \gtrsim 0.1 m_{\chi}$) it is vice versa. Figure \ref{fig2}, right shows an example of the surface brightness distributions for the case of subhalo, which is optimal for detection and resolving in terms of its parameters like mass etc. Here the energy range is limited to 10--100 GeV. Also the approximate point spread function (PSF) of our instrument is shown, as well as an estimate of the brightness of the isotropic background at this energies according to Fermi-LAT data \cite{Fermi}. We did not include the Galactic background since we assumed an optimistic case for detection, i.e. that the subhalo is located at high Galactic latitudes. We see that the subhalo brightness curve lies above the PSF for some angular distance range before fusing with the background. Saying by other words, the convolution of the subhalo brightness distribution with PSF would significantly differ from that in the case of a point source. This in principle allows one to resolve the subhalo inner structure and, hence, to classify such an object unambiguously. However, the total flux from such a subhalo in the mentioned energy range is not higher than $\sim 1$ ph/year, which would not allow to accumulate enough photons by GAMMA-400 alone. If we would go to lower energies, where the spectral flux is higher (see Figure \ref{fig2}, left), this would not help since GAMMA-400 angular resolution worsens with energy decrease \cite{2017EPJWC.14506001T}. The solution of this problem could be in merging the future GAMMA-400 data with data from other telescopes - Fermi-LAT, e-ASTROGAM, etc. - and doing the joint analysis. This might allow one to synthesize the spatial image of a subhalo.

\section{Implications for the subhalo candidates 3FGL J2212.5+0703 and J1924.8--1034}\label{J}

As we already mentioned in Section \ref{intro}, the recent works \cite{2016JCAP...05..049B, 2016PhRvD..94l3002W, 2017PhRvD..95j2001X} confidently reported about a couple of subhalo candidates in 3FGL catalog - J2212.5+0703 and J1924.8--1034. Both of these objects are likely extended with the angular radius $\sim 0.1\degree$. Their spectra can be well fitted by WIMP with mass $\sim$(20--40) GeV annihilating to $b \bar{b}$. The authors of \cite{2016PhRvD..94l3002W, 2017PhRvD..95j2001X} assumed the thermal annihilation cross section $\langle \sigma v \rangle \approx 3 \cdot 10^{-26}$ cm$^3$/s when fitted the signals from candidates. We would like to notice that such a big cross section is unrealistic for the mentioned WIMP masses according to the Fermi-LAT constraints \cite{2015PhRvL.115w1301A} - they require the cross section to be by several times smaller. We re-calculated the required $J$-factors (which reproduce the fluxes) for these objects with the realistic cross section values and obtained $J \geqslant 5 \cdot 10^{20}$ GeV$^2$ cm$^{-5}$. All the known dwarfs have $J \lesssim 10^{20}$ GeV$^2$ cm$^{-5}$ \cite{2017ApJ...834..110A}. Therefore, the subhalo candidates require abnormally big $J$. It seems unlikely for such objects to not be bright and prominent dwarfs. One opportunity to stay dark for them is to be light, but extremely close subhalos. In this context, we estimated the required distances to these objects based on the measured flux from them. We assumed two representative mass values for the subhalo candidates -- $10^6 M_{\odot}$ and $10^9 M_{\odot}$. For these cases we obtained the distances $\sim 1$ kpc and $\sim 10$ kpc respectively (approximately same for both objects). This clearly contradicts the relation (\ref{eq-d}). Also one may have a look at Figure 10, right in \cite{PhysRevD.96.063009}, where the authors plotted all the subhalos from 100 Monte Carlo simulations of MW. And one can clearly see that no single subhalo out of 100 realizations felt closer than $\approx 12$ kpc. Thus, these objects essentially fail both basic tests - by $J$-factor and by distance to them. Indeed, in principle we might be extremely lucky to deal with super big outliers from average trends, which appeared so close to us by a random chance. However, considering Figure 10 in \cite{PhysRevD.96.063009} and the fact that these are two objects - not just one - such a coincidence does not seem to be possible. Instead, these objects are likely typical astrophysical sources composed by two objects projected onto very close lines of sight. This opportunity is absolutely allowed by \cite{2016JCAP...05..049B, 2016PhRvD..94l3002W, 2017PhRvD..95j2001X}. In fact, the data even slightly prefers the binary object hypothesis, since the object images seem to have a mild decline from the circular shape.

These subhalo candidates would be relatively bright sources for GAMMA-400 with the fluxes $\sim 1000$ ph/year above 100 MeV. Indeed, we may include them in our program of observations to study them more. However, based on the considerations above, we do not suppose that these objects are real DM subhalos.

\section{Conclusions and discussion}\label{concl}

In this work we evaluated the detectability of the Galactic DM subhalos by the planned GAMMA-400 gamma-ray telescope. We focused on subhalos with the most relevant masses $(10^6-10^9)M_{\odot}$. We modeled the distribution of DM density inside the subhalos and the spatial distribution of subhalos in MW based on the results of respective N-body simulations of the MW-like halo \cite{PhysRevD.96.063009}. Then we calculated the gamma-ray fluxes from subhalos due to DM annihilation and studied the flux dependence on various parameters (see Figures \ref{fig1},\ref{fig2}). We may outline the following main conclusions.
\begin{itemize}
	\item The total flux (above 100 MeV) from the subhalos rather mildly depends on their mass, but strongly depends on the assumed subhalo concentration and DM particle mass. The most massive subhalos are the brightest despite being the remotest.
	\item For the case of the most relevant WIMP mass $\sim 100$ GeV (and thermal annihilation cross section) we may expect $\gtrsim 1$ objects on the sky with fluxes $\sim 10$ ph/year. This suggests that the subhalos are marginally detectable by GAMMA-400 even for the case of a quite heavy WIMP.
	\item It is difficult to resolve the inner structure of subhalos due to smallness of fluxes. However, this difficulty might be alleviated in the future by the joint analysis of GAMMA-400 data and data from other telescopes - Fermi-LAT, e-ASTROGAM, etc.
	\item The recent subhalo candidates 3FGL J2212.5+0703 and J1924.8--1034 require unrealistically large $J$-factors to reproduce their observed fluxes. We suppose that it is much more probable for these sources to be composed of binary astrophysical sources instead of annihilating DM.
\end{itemize}

Thus, DM subhalos are interesting and promising targets for DM searches by the GAMMA-400 gamma-ray telescope. We plan to continue the research of these objects. Also we plan to study the opportunities to detect and constrain DM in Milky Way satellites by GAMMA-400. This study is supported by the Space Council of the Russian Academy of Sciences and Roskosmos.


\bibliography{C:/Users/Andrey/YandexDisk/DM/universal}

\begin{thebibliography}{20}
\expandafter\ifx\csname natexlab\endcsname\relax\def\natexlab#1{#1}\fi
\expandafter\ifx\csname bibnamefont\endcsname\relax
  \def\bibnamefont#1{#1}\fi
\expandafter\ifx\csname bibfnamefont\endcsname\relax
  \def\bibfnamefont#1{#1}\fi
\expandafter\ifx\csname citenamefont\endcsname\relax
  \def\citenamefont#1{#1}\fi
\expandafter\ifx\csname url\endcsname\relax
  \def\url#1{\texttt{#1}}\fi
\expandafter\ifx\csname urlprefix\endcsname\relax\def\urlprefix{URL }\fi
\providecommand{\bibinfo}[2]{#2}
\providecommand{\eprint}[2][]{\url{#2}}

\bibitem[{\citenamefont{Bertone}(2010)}]{Bertone-book}

\refitem{book}
\bibinfo{author}{\bibfnamefont{G.}~\bibnamefont{Bertone}},
  \emph{\bibinfo{title}{Particle dark matter - observations, models and
  searches}} (\bibinfo{publisher}{Cambridge University Press},
  \bibinfo{year}{2010}).

\bibitem[{\citenamefont{{Charles} \emph{et~al.}}(2016)\citenamefont{{Charles},
  {S{\'a}nchez-Conde}, {Anderson}, {Caputo}, {Cuoco}
  \emph{et~al.}}}]{2016PhR...636....1C}

\refitem{article}
\bibinfo{author}{\bibfnamefont{E.}~\bibnamefont{{Charles}}},
  \bibinfo{author}{\bibfnamefont{M.}~\bibnamefont{{S{\'a}nchez-Conde}}},
  \bibinfo{author}{\bibfnamefont{B.}~\bibnamefont{{Anderson}}},
  \bibinfo{author}{\bibfnamefont{R.}~\bibnamefont{{Caputo}}},
  \bibinfo{author}{\bibfnamefont{A.}~\bibnamefont{{Cuoco}}},
  \bibnamefont{\emph{et~al.}}, \bibinfo{journal}{\physrep}
  \textbf{\bibinfo{volume}{636}}, \bibinfo{pages}{1} (\bibinfo{year}{2016}),
  \eprint{1605.02016}.

\bibitem[{\citenamefont{Calore \emph{et~al.}}(2017)\citenamefont{Calore,
  De~Romeri, Di~Mauro, Donato, and Marinacci}}]{PhysRevD.96.063009}

\refitem{article}
\bibinfo{author}{\bibfnamefont{F.}~\bibnamefont{Calore}},
  \bibinfo{author}{\bibfnamefont{V.}~\bibnamefont{De~Romeri}},
  \bibinfo{author}{\bibfnamefont{M.}~\bibnamefont{Di~Mauro}},
  \bibinfo{author}{\bibfnamefont{F.}~\bibnamefont{Donato}}, \bibnamefont{and}
  \bibinfo{author}{\bibfnamefont{F.}~\bibnamefont{Marinacci}},
  \bibinfo{journal}{Phys. Rev. D} \textbf{\bibinfo{volume}{96}},
  \bibinfo{pages}{063009} (\bibinfo{year}{2017}), \eprint{1611.03503}.

\bibitem[{\citenamefont{{Acero} \emph{et~al.}}(2015)\citenamefont{{Acero},
  {Ackermann}, {Ajello}, {Albert}, {Atwood}
  \emph{et~al.}}}]{2015ApJS..218...23A}

\refitem{article}
\bibinfo{author}{\bibfnamefont{F.}~\bibnamefont{{Acero}}},
  \bibinfo{author}{\bibfnamefont{M.}~\bibnamefont{{Ackermann}}},
  \bibinfo{author}{\bibfnamefont{M.}~\bibnamefont{{Ajello}}},
  \bibinfo{author}{\bibfnamefont{A.}~\bibnamefont{{Albert}}},
  \bibinfo{author}{\bibfnamefont{W.~B.} \bibnamefont{{Atwood}}},
  \bibnamefont{\emph{et~al.}}, \bibinfo{journal}{\apjs}
  \textbf{\bibinfo{volume}{218}}, \bibinfo{eid}{23} (\bibinfo{year}{2015}),
  \eprint{1501.02003}.

\bibitem[{\citenamefont{{Bertoni} \emph{et~al.}}(2016)\citenamefont{{Bertoni},
  {Hooper}, and {Linden}}}]{2016JCAP...05..049B}

\refitem{article}
\bibinfo{author}{\bibfnamefont{B.}~\bibnamefont{{Bertoni}}},
  \bibinfo{author}{\bibfnamefont{D.}~\bibnamefont{{Hooper}}}, \bibnamefont{and}
  \bibinfo{author}{\bibfnamefont{T.}~\bibnamefont{{Linden}}},
  \bibinfo{journal}{\jcap} \textbf{\bibinfo{volume}{5}}, \bibinfo{eid}{049}
  (\bibinfo{year}{2016}), \eprint{1602.07303}.

\bibitem[{\citenamefont{{Wang} \emph{et~al.}}(2016)\citenamefont{{Wang},
  {Duan}, {Ma}, {Liang}, {Shen} \emph{et~al.}}}]{2016PhRvD..94l3002W}

\refitem{article}
\bibinfo{author}{\bibfnamefont{Y.-P.} \bibnamefont{{Wang}}},
  \bibinfo{author}{\bibfnamefont{K.-K.} \bibnamefont{{Duan}}},
  \bibinfo{author}{\bibfnamefont{P.-X.} \bibnamefont{{Ma}}},
  \bibinfo{author}{\bibfnamefont{Y.-F.} \bibnamefont{{Liang}}},
  \bibinfo{author}{\bibfnamefont{Z.-Q.} \bibnamefont{{Shen}}},
  \bibnamefont{\emph{et~al.}}, \bibinfo{journal}{\prd}
  \textbf{\bibinfo{volume}{94}}, \bibinfo{eid}{123002} (\bibinfo{year}{2016}),
  \eprint{1611.05135}.

\bibitem[{\citenamefont{{Xia} \emph{et~al.}}(2017)\citenamefont{{Xia}, {Duan},
  {Li}, {Liang}, {Shen} \emph{et~al.}}}]{2017PhRvD..95j2001X}

\refitem{article}
\bibinfo{author}{\bibfnamefont{Z.-Q.} \bibnamefont{{Xia}}},
  \bibinfo{author}{\bibfnamefont{K.-K.} \bibnamefont{{Duan}}},
  \bibinfo{author}{\bibfnamefont{S.}~\bibnamefont{{Li}}},
  \bibinfo{author}{\bibfnamefont{Y.-F.} \bibnamefont{{Liang}}},
  \bibinfo{author}{\bibfnamefont{Z.-Q.} \bibnamefont{{Shen}}},
  \bibnamefont{\emph{et~al.}}, \bibinfo{journal}{\prd}
  \textbf{\bibinfo{volume}{95}}, \bibinfo{eid}{102001} (\bibinfo{year}{2017}),
  \eprint{1611.05565}.

\bibitem[{\citenamefont{{Chou} \emph{et~al.}}(2017)\citenamefont{{Chou},
  {Tanoglidis}, and {Hooper}}}]{2017arXiv170908562C}

\refitem{article}
\bibinfo{author}{\bibfnamefont{T.-L.} \bibnamefont{{Chou}}},
  \bibinfo{author}{\bibfnamefont{D.}~\bibnamefont{{Tanoglidis}}},
  \bibnamefont{and} \bibinfo{author}{\bibfnamefont{D.}~\bibnamefont{{Hooper}}},
  \bibinfo{journal}{ArXiv e-prints}  (\bibinfo{year}{2017}),
  \eprint{1709.08562}.

\bibitem[{\citenamefont{{Galper} \emph{et~al.}}(2013)\citenamefont{{Galper},
  {Adriani}, {Aptekar}, {Arkhangelskaja}, {Arkhangelskiy}
  \emph{et~al.}}}]{2013AdSpR..51..297G}

\refitem{article}
\bibinfo{author}{\bibfnamefont{A.~M.} \bibnamefont{{Galper}}},
  \bibinfo{author}{\bibfnamefont{O.}~\bibnamefont{{Adriani}}},
  \bibinfo{author}{\bibfnamefont{R.~L.} \bibnamefont{{Aptekar}}},
  \bibinfo{author}{\bibfnamefont{I.~V.} \bibnamefont{{Arkhangelskaja}}},
  \bibinfo{author}{\bibfnamefont{A.~I.} \bibnamefont{{Arkhangelskiy}}},
  \bibnamefont{\emph{et~al.}}, \bibinfo{journal}{Advances in Space Research}
  \textbf{\bibinfo{volume}{51}}, \bibinfo{pages}{297} (\bibinfo{year}{2013}),
  \eprint{1201.2490}.

\bibitem[{\citenamefont{{Topchiev}
  \emph{et~al.}}(2017)\citenamefont{{Topchiev}, {Galper}, {Bonvicini},
  {Adriani}, {Arkhangelskaja} \emph{et~al.}}}]{2017EPJWC.14506001T}

\refitem{inproceedings}
\bibinfo{author}{\bibfnamefont{N.~P.} \bibnamefont{{Topchiev}}},
  \bibinfo{author}{\bibfnamefont{A.~M.} \bibnamefont{{Galper}}},
  \bibinfo{author}{\bibfnamefont{V.}~\bibnamefont{{Bonvicini}}},
  \bibinfo{author}{\bibfnamefont{O.}~\bibnamefont{{Adriani}}},
  \bibinfo{author}{\bibfnamefont{I.~V.} \bibnamefont{{Arkhangelskaja}}},
  \bibnamefont{\emph{et~al.}}, in \emph{\bibinfo{booktitle}{European Physical
  Journal Web of Conferences}} (\bibinfo{year}{2017}), vol.
  \bibinfo{volume}{145}, p. \bibinfo{pages}{06001}.

\bibitem[{\citenamefont{{De Angelis} \emph{et~al.}}(2017)\citenamefont{{De
  Angelis}, {Tatischeff}, {Tavani}, {Oberlack}, {Grenier}
  \emph{et~al.}}}]{2017ExA...tmp...24D}

\refitem{article}
\bibinfo{author}{\bibfnamefont{A.}~\bibnamefont{{De Angelis}}},
  \bibinfo{author}{\bibfnamefont{V.}~\bibnamefont{{Tatischeff}}},
  \bibinfo{author}{\bibfnamefont{M.}~\bibnamefont{{Tavani}}},
  \bibinfo{author}{\bibfnamefont{U.}~\bibnamefont{{Oberlack}}},
  \bibinfo{author}{\bibfnamefont{I.}~\bibnamefont{{Grenier}}},
  \bibnamefont{\emph{et~al.}}, \bibinfo{journal}{Experimental Astronomy}
  \textbf{\bibinfo{volume}{44}}, \bibinfo{pages}{25} (\bibinfo{year}{2017}),
  \eprint{1611.02232}.

\bibitem[{\citenamefont{{Martinez}}(2015)}]{2015MNRAS.451.2524M}

\refitem{article}
\bibinfo{author}{\bibfnamefont{G.~D.} \bibnamefont{{Martinez}}},
  \bibinfo{journal}{\mnras} \textbf{\bibinfo{volume}{451}},
  \bibinfo{pages}{2524} (\bibinfo{year}{2015}), \eprint{1309.2641}.

\bibitem[{\citenamefont{{Geringer-Sameth}
  \emph{et~al.}}(2015)\citenamefont{{Geringer-Sameth}, {Koushiappas}, and
  {Walker}}}]{2015ApJ...801...74G}

\refitem{article}
\bibinfo{author}{\bibfnamefont{A.}~\bibnamefont{{Geringer-Sameth}}},
  \bibinfo{author}{\bibfnamefont{S.~M.} \bibnamefont{{Koushiappas}}},
  \bibnamefont{and} \bibinfo{author}{\bibfnamefont{M.}~\bibnamefont{{Walker}}},
  \bibinfo{journal}{\apj} \textbf{\bibinfo{volume}{801}}, \bibinfo{eid}{74}
  (\bibinfo{year}{2015}), \eprint{1408.0002}.

\bibitem[{PPP()}]{PPPC}

\refitem{misc}
\bibinfo{note}{\url{http://www.marcocirelli.net/PPPC4DMID.html}}.

\bibitem[{\citenamefont{{Cirelli} \emph{et~al.}}(2011)\citenamefont{{Cirelli},
  {Corcella}, {Hektor}, {H{\"u}tsi}, {Kadastik}
  \emph{et~al.}}}]{2011JCAP...03..051C}

\refitem{article}
\bibinfo{author}{\bibfnamefont{M.}~\bibnamefont{{Cirelli}}},
  \bibinfo{author}{\bibfnamefont{G.}~\bibnamefont{{Corcella}}},
  \bibinfo{author}{\bibfnamefont{A.}~\bibnamefont{{Hektor}}},
  \bibinfo{author}{\bibfnamefont{G.}~\bibnamefont{{H{\"u}tsi}}},
  \bibinfo{author}{\bibfnamefont{M.}~\bibnamefont{{Kadastik}}},
  \bibnamefont{\emph{et~al.}}, \bibinfo{journal}{\jcap}
  \textbf{\bibinfo{volume}{3}}, \bibinfo{eid}{051} (\bibinfo{year}{2011}),
  \eprint{1012.4515}.

\bibitem[{\citenamefont{{Ciafaloni}
  \emph{et~al.}}(2011)\citenamefont{{Ciafaloni}, {Comelli}, {Riotto}, {Sala},
  {Strumia} \emph{et~al.}}}]{2011JCAP...03..019C}

\refitem{article}
\bibinfo{author}{\bibfnamefont{P.}~\bibnamefont{{Ciafaloni}}},
  \bibinfo{author}{\bibfnamefont{D.}~\bibnamefont{{Comelli}}},
  \bibinfo{author}{\bibfnamefont{A.}~\bibnamefont{{Riotto}}},
  \bibinfo{author}{\bibfnamefont{F.}~\bibnamefont{{Sala}}},
  \bibinfo{author}{\bibfnamefont{A.}~\bibnamefont{{Strumia}}},
  \bibnamefont{\emph{et~al.}}, \bibinfo{journal}{\jcap}
  \textbf{\bibinfo{volume}{3}}, \bibinfo{eid}{019} (\bibinfo{year}{2011}),
  \eprint{1009.0224}.

\bibitem[{\citenamefont{{Molin{\'e}}
  \emph{et~al.}}(2017)\citenamefont{{Molin{\'e}}, {S{\'a}nchez-Conde},
  {Palomares-Ruiz}, and {Prada}}}]{2017MNRAS.466.4974M}

\refitem{article}
\bibinfo{author}{\bibfnamefont{{\'A}.}~\bibnamefont{{Molin{\'e}}}},
  \bibinfo{author}{\bibfnamefont{M.~A.} \bibnamefont{{S{\'a}nchez-Conde}}},
  \bibinfo{author}{\bibfnamefont{S.}~\bibnamefont{{Palomares-Ruiz}}},
  \bibnamefont{and} \bibinfo{author}{\bibfnamefont{F.}~\bibnamefont{{Prada}}},
  \bibinfo{journal}{\mnras} \textbf{\bibinfo{volume}{466}},
  \bibinfo{pages}{4974} (\bibinfo{year}{2017}), \eprint{1603.04057}.

\bibitem[{\citenamefont{{Ackermann}
  \emph{et~al.}}(2015)\citenamefont{{Ackermann}, {Albert}, {Anderson},
  {Atwood}, {Baldini} \emph{et~al.}}}]{2015PhRvL.115w1301A}

\refitem{article}
\bibinfo{author}{\bibfnamefont{M.}~\bibnamefont{{Ackermann}}},
  \bibinfo{author}{\bibfnamefont{A.}~\bibnamefont{{Albert}}},
  \bibinfo{author}{\bibfnamefont{B.}~\bibnamefont{{Anderson}}},
  \bibinfo{author}{\bibfnamefont{W.~B.} \bibnamefont{{Atwood}}},
  \bibinfo{author}{\bibfnamefont{L.}~\bibnamefont{{Baldini}}},
  \bibnamefont{\emph{et~al.}}, \bibinfo{journal}{Physical Review Letters}
  \textbf{\bibinfo{volume}{115}}, \bibinfo{eid}{231301} (\bibinfo{year}{2015}),
  \eprint{1503.02641}.

\bibitem[{Fer()}]{Fermi}

\refitem{misc}
\bibinfo{note}{\url{https://fermi.gsfc.nasa.gov/ssc/data/access/}}.

\bibitem[{\citenamefont{{Albert} \emph{et~al.}}(2017)\citenamefont{{Albert},
  {Anderson}, {Bechtol}, {Drlica-Wagner}, {Meyer}
  \emph{et~al.}}}]{2017ApJ...834..110A}

\refitem{article}
\bibinfo{author}{\bibfnamefont{A.}~\bibnamefont{{Albert}}},
  \bibinfo{author}{\bibfnamefont{B.}~\bibnamefont{{Anderson}}},
  \bibinfo{author}{\bibfnamefont{K.}~\bibnamefont{{Bechtol}}},
  \bibinfo{author}{\bibfnamefont{A.}~\bibnamefont{{Drlica-Wagner}}},
  \bibinfo{author}{\bibfnamefont{M.}~\bibnamefont{{Meyer}}},
  \bibnamefont{\emph{et~al.}}, \bibinfo{journal}{\apj}
  \textbf{\bibinfo{volume}{834}}, \bibinfo{eid}{110} (\bibinfo{year}{2017}),
  \eprint{1611.03184}.

\end{thebibliography}
%
%
%
%

\end{document}